\documentclass[11pt]{article}
\usepackage[utf8]{inputenc}

\title{Solvent-dependent termination, size and stability in polyynes synthesis by laser ablation in liquids
}
\date{}

\usepackage[super,sort&compress,comma]{natbib}
\usepackage{balance}
\usepackage{mathptmx}
\usepackage{sectsty}
\usepackage{graphicx} 
\usepackage{lastpage}
\usepackage{float}
\usepackage{amsmath}
\usepackage{fancyhdr}
\usepackage{fnpos}
\usepackage[left=1.5cm, right=1.5cm, top=2.0cm, bottom=2.0cm]{geometry}
\usepackage[english]{babel}
\usepackage{array}
\usepackage{droidsans}
\usepackage{charter}
\usepackage[T1]{fontenc}
\usepackage{setspace}
\usepackage[compact]{titlesec}
\usepackage{hyperref}
\usepackage{amssymb}
\usepackage{authblk}
\usepackage{pdfpages}

\author[1]{Sonia~Peggiani}
\author[1]{Pietro~Marabotti}
\author[1]{Riccardo~Alberto~Lotti}
\author[1]{Anna~Facibeni}
\author[1]{Patrick~Serafini}
\author[1]{Alberto~Milani}
\author[1]{Valeria~Russo}
\author[1]{Andrea~Li~Bassi}
\author[1]{Carlo~Spartaco~Casari}
\affil[1]{Department of Energy, Politecnico di Milano, Via Ponzio 34/3, 20133, Milano, Italy}

\begin{document}

\maketitle

In recent years there has been a growing interest in sp-carbon chains as possible novel nanostructures. An example of sp-carbon chains are the so-called polyynes, characterized by the alternation of single and triple bonds that can be synthesized by pulsed laser ablation in liquid (PLAL) of a graphite target. In this work, by exploiting different solvents in the PLAL process, e.g. water, acetonitrile, methanol, ethanol, and isopropanol, we systematically investigate the solvent role in polyyne formation and stability. The presence of methyl- and cyano- groups in the solutions influences the termination of polyynes, allowing to detect, in addition to hydrogen-capped polyynes up to HC$_{22}$H, methyl-capped polyynes up to 18 carbon atoms (i.e. HC$_n$CH$_3$) and cyanopolyynes up to HC$_{12}$CN. The assignment of each species was done by UV-Vis spectroscopy and supported by density functional theory simulations of vibronic spectra. In addition, surface-enhanced Raman spectroscopy allowed to observe differences, due to different terminations (hydrogen, methyl-and cyano group), in the shape and positions of the characteristic Raman bands of the size-selected polyynes. The evolution in time of each polyyne has been investigated evaluating the chromatographic peak area, and the effect of size, terminations and solvents on polyynes stability has been individuated.
\section{Introduction}
Polyynes are carbon nanostructures, consisting in finite linear chains of sp-carbon atoms linked by alternated single and triple bonds that show appealing tunability of optical and electronic properties by controlling the length of the chains and their terminations \cite{casari_milani_2018,casari2016}. Among the large variety of available chemical and physical methods to synthesize sp-carbon chains \cite{eastmond1972,cataldo2003,chalifoux2010}, pulsed laser ablation in liquid (PLAL) is able to produce isolated polyynes with different lengths in an easy and efficient way \cite{taguchi2015}. PLAL is based on the irradiation of a carbon target in a solvent by means of laser pulses \cite{tsuji2002,tsuji2003}. During the last years, many research groups intended to synthetize polyynes by PLAL have employed ns-laser pulses. A few works based on ablation in water reported the synthesis of H-polyynes only up to H-C$_{12}$-H (for simplicity here called C$_{12}$) \cite{compagnini2007,forte2013,grasso2009}. Matsutani \textit{et al.}, instead, presented the ablation of PCDTA and graphite pellet in different alcohols (methanol, ethanol, 1-propanol, 1-butanol, t-butyl alcohol) at 532 nm, obtaining polyynes length of 18 atoms \cite{matsutani2011,matsutani2008}. They also reported the formation of C$_{22}$ by irradiating graphite and fullerene suspension in n-hexane at laser wavelength of 532 nm. The longest hydrogen-capped polyynes ever obtained by PLAL, C$_{30}$ , has been obtained employing decalin at 1064 nm \cite{matsutani2012,matsutani2011_2}. In addition to H-polyynes, other two sequences of sp-carbon chains detected by ablating graphite target in acetonitrile were cyanopolyynes, i.e. H-C$_n$-CN (n=6,8,10,12), reported in the paper of Wakabayashi \textit{et al.}\cite{wakabayashi2012} and dicyanopolyynes, C$_n$N$_2$ (n=6,8), found by Forte \textit{et al.}\cite{forte2013}. Moreover, polyynes capped by methyl-group, i.e. H-C$_n$-CH$_3$ (n=8,10,12), were synthesized in hexane \cite{Wada2012} and in toluene, employing fs-laser at the meniscus of the liquid \cite{ramadhan2017}. Nevertheless, no systematic studies of the effect of different solvents on the properties of polyynes, e.g. yield, size and terminations, during PLAL experiments have been performed at the same fluence, wavelength and liquid volume (see \textbf{Table S.1.} in the Supporting Information). \\
The solutions obtained after the ablation have been commonly studied by UV-Vis spectroscopy from which polyynes concentration can be extracted \cite{taguchi2015,taguchi2017}. High-performance liquid chromatography (HPLC) can be employed to separate in lengths and terminations the sp-carbon chains, allowing the characterization and collection of size-selected polyynes \cite{cataldo2005}. Moreover, Raman spectroscopy allows to recognize sp-hybridized linear carbon chains because of their CC stretching mode in the frequency range of 1800-2200 cm$^{-1}$, where no peaks related to other functional groups are present \cite{kurti1995,heiman1999}. Simulated and experimental Raman spectra show redshifts in correspondence of longer chain length, due to the increase of the conjugation \cite{Lucotti2006,Tabata2006}. Weak or undetectable Raman signals can be due to low concentrated samples or to a strong luminescence signals, probably linked to the by-products of PLAL \cite{Tabata2006}. In this case, it is possible to exploit noble metal nanoparticles to enhance the sensitivity of the Raman signals up to six orders of magnitude carrying out surface-enhanced Raman spectroscopy (SERS) measurements \cite{Lucotti2006}. Despite the interest in analysing polyynes with different end-caps, no one to the best of our knowledge examined SERS of size-selected polyynes with terminations different from hydrogen from physical synthesis methods.\\
One of the major issues of polyynes is their poor stability, as they tend to rearrange in more stable sp$^2$-based structures by crosslinking reactions \cite{Casari2004}. Only few works discussed the evolution in time of polyynes in liquid solutions without silver nanoparticles colloids, the study of Cataldo after submerged arc discharge in acetonitrile \cite{Cataldo2007} and the one of Compagnini and Scalese using water during PLAL \cite{Compagnini2012}. Thus, to exploit sp-carbon chains in future applications, there is the need to understand how to obtain stable structures by physical method such as PLAL. In this framework some still open questions regard the solvent effect in determining  the polyne produciton yield, in providing specific terminating functional groups and in affecting the stability of polyynes in liquid. \\
We here provide a systematic investigation of the synthesis of polyynes by PLAL in different solvents such as water, acetonitrile, ethanol, methanol and isopropanol to outline their role in providing terminations for polyynes. UV-Vis spectroscopy, HPLC and theoretical simulations have allowed us to recognize polyynes with length and terminations never reported so far in those solvents (e.g. C$_{22}$ in organic solvents and methylpolyynes in all the solvents). SERS spectra have been obtained from size-selected and termination-selected polyynes. The specific stability of each type of polyyne has been investigated by analysing time constants for different lengths, terminations and solutions. Different solvent properties including polarity, composition, viscosity are shown to compete in polyyne formation and behaviour, thus underlying the complex role of the solvent and the peculiar physical-chemistry of polyynes in liquid environments.

\section{Experimental details}\label{sec2}
Polyynes were synthesized by pulsed laser ablation in liquid (PLAL) of a graphite target with purity of 99.99\% (Testbourne Ltd.), exploiting pulses from the second harmonic ($\lambda$ = 532 nm) of a Nd:YAG pulsed laser (Quantel Q-smart 850) . The laser operated at a frequency of 10 Hz with pulse duration of 6 ns. Ablation was performed at a fixed time of 15 minutes in 2 mL of different solvents: deionised water Milli-Q (0.055 $\mu$S), acetonitrile (ACN), isopropanol (IPA), methanol (MeOH) and ethanol (EtOH) (Sigma-Aldrich, purity $\geq$ 99.9\%). The laser beam was focused on the target with a lens with  200 mm focal length with a fluence of 2.8 J/cm$^2$ for all solvents (the effect of the liquid layer was taken into account in the calculations of the fluence \cite{MenndezManjn2011}). During ablation the sample holder stage was moved along a spiral path to ensure uniform ablation of the graphite target.\\
The ablated solutions were all filtered through Phenomenex Phenex RC-membrane syringe filters having pore size of 0.45 $\mu$m to remove any sizeable impurities. Polyynes presence and concentration were evaluated by UV-Vis absorption spectroscopy (Shimadzu UV-1800 UV/Visible Scanning Spectrophotometer with 190-1100 nm of spectral range). The optical path of the employed quartz cuvettes is of 10 mm. To avoid signal saturation, samples were diluted in the proportion of 1/6 v/v for ablation in water, whereas 1/60 v/v for the other solvents. Moreover, we analysed the UV-Vis spectra from 340 nm to 210 nm due to the high absorption of the organic solvents below 210 nm \cite{krstulovic1982theory}.\\
Reverse-phase high-performance liquid chromatography (RP-HPLC) was performed by means of a Shimadzu Prominence UFLC, equipped with a photodiode array (DAD) UV-Vis spectrometer, FRC-10A fraction collector and C8 column (Shimadzu Shim-pack HPLC packed column GIST, 250 mm x 4.6 mm, 5 $\mu$m particle size). The detection of size- and termination-selected polyynes was achieved exploiting a gradient elution: in this regime, the mobile phase initially consists in a solution of acetonitrile/water 50/50 v/v that gradually evolves to 90/10 v/v in 18 minutes. The overall analysis time was set to 60 min, with a constant flow rate of 1 mL/min. To estimate the stability of polyynes, we performed HPLC analysis at definite time intervals, until 30 days, and we evaluated their degradation from the area under the chromatographic peaks associated to each polyyne. During the analysis period, the samples were stored into closed glass vials at room temperature in dark. In addition, size-selected polyynes can be selectively collected for SERS investigations whose acetonitrile/water relative concentration depends on the collection time.\\
SERS spectra were collected by Renishaw inVia Raman microscope with a diode-pumped solid-state laser ($\lambda$ = 660 nm). Silver colloids ($10^{-3}$ M) were synthesized following Lee-Meisel method \cite{Lee1982}, showing an absorption peak at 413 nm and a predicted mean diameter of $\sim$41 nm \cite{Paramelle2014}. Silver colloids were added to size-separated polyynes 1/4 v/v to implement SERS analysis in liquid. Laser power was set to 75 mW during all measurements.\\
Density Functional Theory (DFT) simulations were performed on single linear chains using Gaussian09[g09]. Time Dependent Density Functional Theory (TD-DFT) calculations were carried for the prediction of the vibronic spectra at CAM-B3LYP/cc-pVTZ level of theory, while PBE0/cc-pVTZ calculations have been used for the computation of the Raman spectra. Based on TD-DFT calculations, Huang-Rys factors and vibronic spectra have been computed by home-made programs \cite{Yang2017, Peggiani2020}. The computed vibrational frequencies were multiplied by a scale factor of 0.961 in the comparison with experimental spectra \cite{nist}. The reliability of the level of theory here chosen is supported by previous studies about UV-Vis prediction and Raman spectra simulations of different sp-carbon based systems, which turned out to be in a very good agreement with the experimental ones \cite{Peggiani2020, Tommasini2007}.

\section{Results and discussion}
\subsection{Characterization of polyynes structure}
UV-Vis absorption spectra of solutions prepared by pulsed laser ablation of graphite in acetonitrile, isopropanol, ethanol,  methanol and pure water are shown in \textbf{Figure 1a}. All the ablations were performed employing the same experimental parameters (laser fluence and wavelength, ablation time, solvent volume) to evaluate the effect of the solvent in the formation of polyynes. Each spectrum presents several peaks assigned to polyynes of different length. Specifically, in all the spectra the 0-0 band of the $^1\Sigma_u^+ \xleftarrow{} X^1\Sigma_g^+$ transition of hydrogen-capped polyyne can be seen at 225 nm (C$_8$) and, for the organic solvents at 251 nm (C$_{10}$), 275 nm (C$_{12}$), 295 nm (C$_{14}$) and 314 nm (C$_{16}$). The solutions in organic solvents are yellowish with a colour intensity varying from brown yellow of acetonitrile to pale yellow of methanol and fully transparent of water in agreement with the absorption spectra (see inset of \textbf{Figure 1}). The spectra show an unresolved absorption background, related with possible formation of hydrocarbons during the synthesis of polyynes. Such background monotonically decreases with the wavelength, as observed by Taguchi and co-workers \cite{taguchi2015,taguchi2017}. The spectra, after subtraction of the background obtained by fitting the baseline with a decreasing exponential and the polyynes peaks with gaussian curves, are reported in \textbf{Figure 1b}. Polyyne production yield was estimated from the absorbance values extracted from \textbf{Figure 1b} by calculating $\chi_p$ as the ratio between the integrated absorption of the 0-0 band of the $^1\Sigma_u^+ \xleftarrow{} X^1\Sigma_g^+$ of H-polyynes to that of fitted background of hydrocarbons by-products \cite{taguchi2015}. The lowest $\chi_p$ was found in the case of water with a value of 0.0749$\pm$0.0001 and the highest “purity” of polyynes over the other contaminants was observed in methanol at 0.3015$\pm$0.0003, followed by isopropanol, 0.2860$\pm$0.0043, ethanol, 0.2766$\pm$0.0035 and acetonitrile with a value of 0.2057$\pm$0.0005. Values of $\chi_p$ have been plotted as a function of the solvent polarity (see \textbf{Figure 2}), considered 100 the polarity of water, and compared with polyyne concentrations (see \textbf{Table S.2} in the Supporting Information) evaluated from Lambert-Beer law, employing molar extinction coefficients taken from literature \cite{eastmond1972}. We observe that polyynes concentration decreases with increasing the polarity of the solvents in agreement with previous works \cite{matsutani2011,Li2020}, whereas, apart from water, we see an opposite behaviour for $\chi_p$ value, which is larger in correspondence of lower polyynes concentration.\\
\begin{figure}[t!]
    \centering
    \includegraphics[width=0.3\paperwidth]{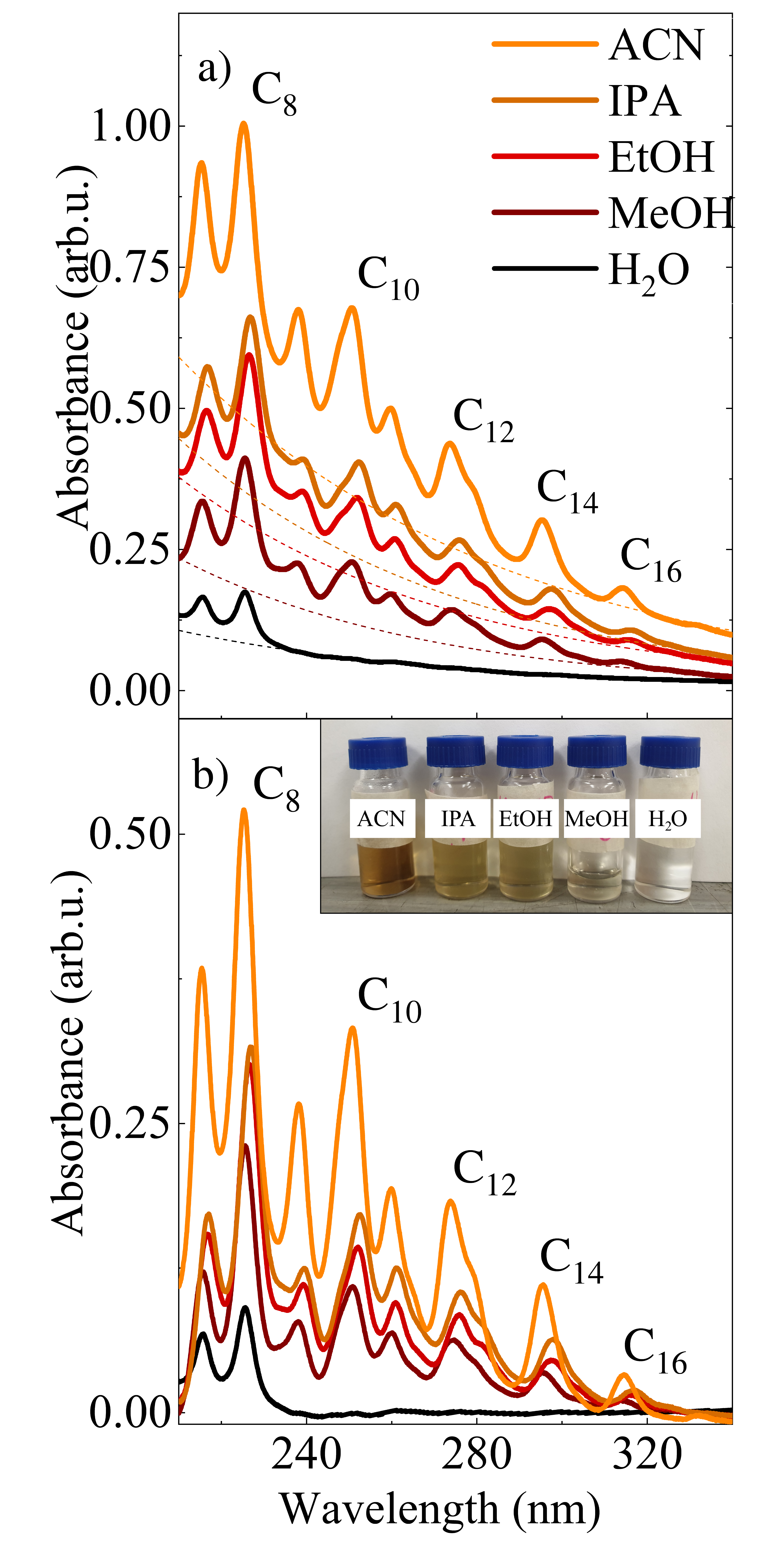}
    \caption{UV-Vis spectra of filtered solutions of polyynes in acetonitrile, isopropanol, methanol, ethanol and water after laser ablation before \textbf{a)} and after \textbf{b)} the background (dotted lines) subtraction. The 0-0 band of the $^1\Sigma_u^+ \xleftarrow{} X_1\Sigma_g^+$ transition of the hydrogen-capped polyynes long 8, 10, 12, 14, 16 atoms of carbon are respectively referred to as C$_8$, C$_{10}$, C$_{12}$, C$_{14}$, C$_{16}$.  Picture of the different solutions is reported in the inset.} 
\end{figure}
\\
Polyynes, which are nonpolar molecules and liable to oxidation \cite{Cataldo2004}, are more concentrated in acetonitrile probably because the solvent has the lowest value of polarity \cite{handbooksolv} and the slightest quantity of oxygen dissolved compared to the other solvents (see \textbf{Table S.2} in the Supporting Information for the values of oxygen present in the other solvents) \cite{Bebahani2002}. Moreover, acetonitrile has a higher tendency towards carbonization with respect to alcohols, in which this process is thermodynamically unfavourable \cite{Cataldo2004_2}. Indeed, alcohols contain oxygen and are characterized by H/C ratios respectively of 0.33 for MeOH, 0.4 for EtOH, 0.5 for IPA, which are lower than that of ACN (i.e. 0.66).  The coking reaction could give a secondary source of carbon species in the plasma phase \cite{wakabayashi2012,Amendola2013} but, in parallel, it causes the formation of hydrocarbons which affects the $\chi_p$ value. For these reasons, the solution of polyynes synthesized in acetonitrile, despite the high concentration, turns out to be characterized by a lower value of $\chi_p$ than alcohols, corresponding to the larger hydrocarbon-correlated background, as mentioned before \cite{Cataldo2004_2}. In case of water, we observe the lowest concentration of polyynes and the smallest value of $\chi_p$, thus exhibiting the worst yield of polyynes. This result can be explained considering that water does not contribute to furnish carbon atoms in the polyyne formaton process \cite{wakabayashi2012} and its hydrogen generation rate, which is an important factor in polyynes formation \cite{tsuji2003,park2012}, is the lowest one with respect to the organic solvents. In fact, the H–OH bond in water (4.8 eV) is less likely to break than the molecular bonds in organic liquids, such as C–H (4.3 eV), C–C (3.6 eV) and C–O (3.7 eV) \cite{handbookchem,Kanitz2019}.
\begin{figure}[th]
    \centering
    \includegraphics[width=0.4\paperwidth]{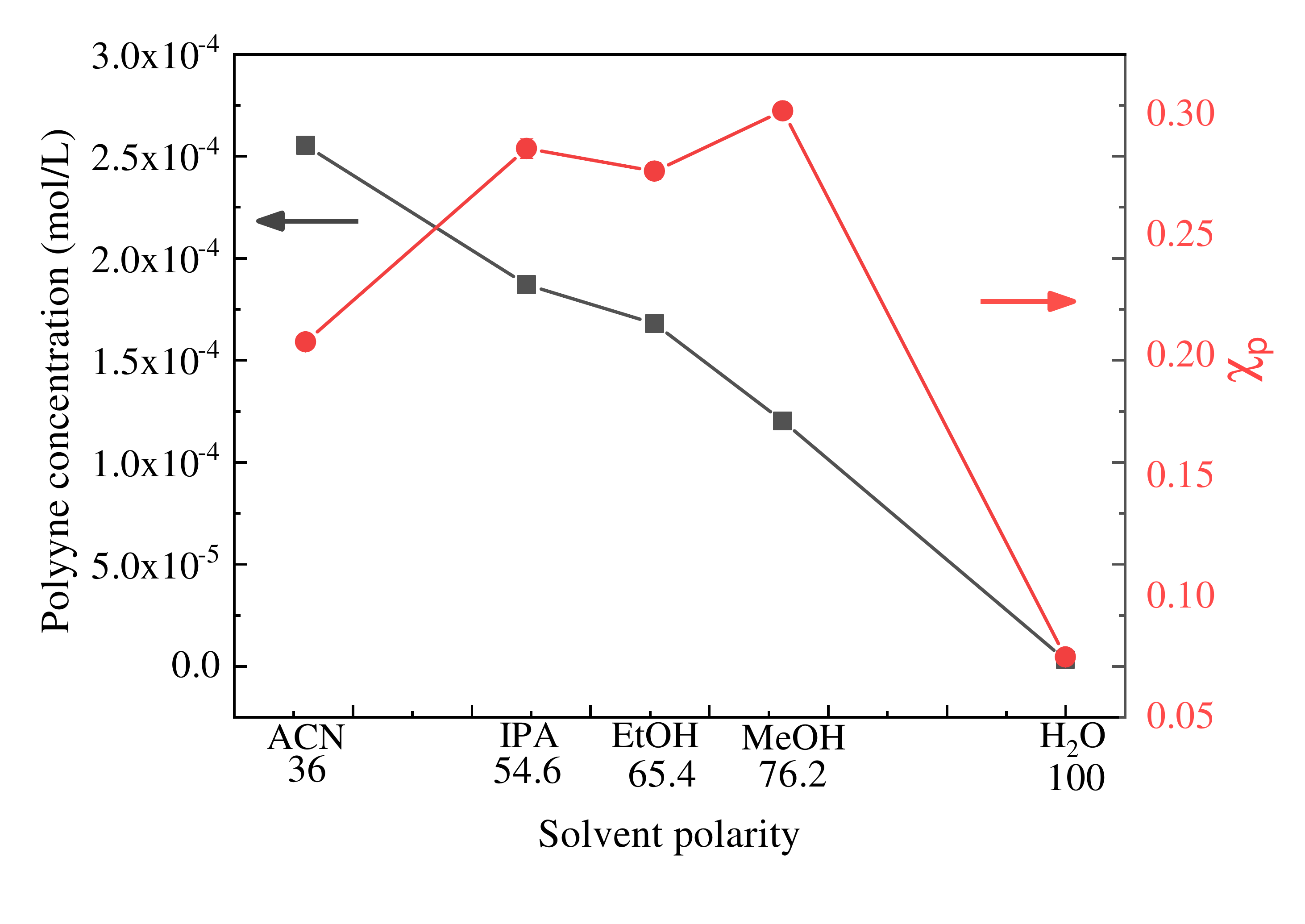}
    \caption{Overall polyyne concentration (mol/L) and $\chi_p$ as a function of the solvent polarity.}
\end{figure}
\\HPLC allows for a detailed investigation of the length and termination of the polyynes present in the samples. After processing the data obtained from the DAD coupled with HPLC, a 2D graph of the absorbance as a function of time and wavelength shows the presence of different polyyne species eluted between 10 and 32 minutes in the acetonitrile solution (see \textbf{Figure 3a}). All the corresponding times on the chromatogram and the positions of the experimental, simulated and literature UV-Vis absorption peaks are reported in \textbf{Table S.3} in the Supporting Information, together with related references of previous investigations \cite{matsutani2012,wakabayashi2012,Wada2012,ramadhan2017,Peggiani2020}. We detected nine different species of hydrogen-capped polyynes ranging from 6 to 22 sp-carbon atoms, whose UV-Vis spectra are reported in \textbf{Figure 3b}. The same size of H-polyynes was found also in the case of methanol, ethanol and isopropanol at analogous times as those in acetonitrile. The species C$_6$ was not discussed previously because its UV-Vis absorption peaks fall in the UV cut-off region of the alcohols. Moreover, we individuated other types of sp-carbon wires, which we attributed to polyynes terminated with methyl (H-C$_n$-CH$_3$) and cyano (H-C$_n$-CN) groups thanks to a multi-technique characterization based on experimental and computed data as discussed later in this section. The corresponding UV-Vis spectra are represented in \textbf{Figure 3c} and \textbf{3d}, respectively. Cyanopolyynes were obtained only by laser ablation in acetonitrile, while methylpolyynes were noticed also by ablation in all the other solvents, included water, where it was only detected the species HC$_8$CH$_3$. The development of an optimized HPLC separation method (see Section \ref{sec2}), able to differentiate two very similar constituents, allowed to distinguish cyano- and methyl- from H-polyynes. To confirm our assignments, we first considered the times at which the species were eluted by the chromatographic system (see \textbf{Figure 3a} and \textbf{Table S.3} in the Supporting Information). Each peak appeared on the chromatogram according to the size of the polyyne molecules from the shorter to the longest. We observed between the peaks of two neighbouring hydrogen-capped polyynes the presence of cyanopolyynes followed by methylpolyynes, which both have an odd number of carbon atoms. When molecules are characterized by analogous dimension, the driven factor for their separation is their polarity: higher polar molecules are less retained by the HPLC column \cite{hplc}. For this reason, cyanopolyynes, which are more polar than methylpolyynes, were eluted first, as showed in \textbf{Figure 3a)}. Secondly, to support the interpretation of our UV-Vis data, we performed TD-DFT simulations to compute vibronic spectra. The predicted data were rigidly shifted to match the spectra of C$_8$, considered as a reference system as in our previous work \cite{Peggiani2020}. Transition energies are well described by the TD-DFT model, even if by increasing the length of the chain, the discrepancy with respect to the experimental values increases. This effect can be due to partial multireference behaviour of the polyynes excited states not considered in TD-DFT \cite{Peggiani2020}. Finally, we observe good agreement between the experimental main UV absorption peaks and those reported in literature, except for small shifts of approximately 2-3 nm in some cases due to the specific liquid environment \cite{absorption_book}. Similar chain lengths were reported in previous works for hydrogen-capped polyynes by ablation in solvents as hexane, n-hexane and decalin  \cite{matsutani2012,matsutani2011_2,Matsutani2009,Inoue2010} and for cyanopolyynes by ablation in acetonitrile \cite{forte2013,wakabayashi2012} while only methyl-capped polyynes with n= 10, 12, 14 were discussed in literature \cite{Wada2012,ramadhan2017}. Our results suggest the formation of other methyl-capped polyynes with n=6, 8, 16, 18 and show first-time detection of a methyl-capped polyyne in water.\\
\begin{figure*}[t!]
    \centering
    \includegraphics[width=0.6\paperwidth]{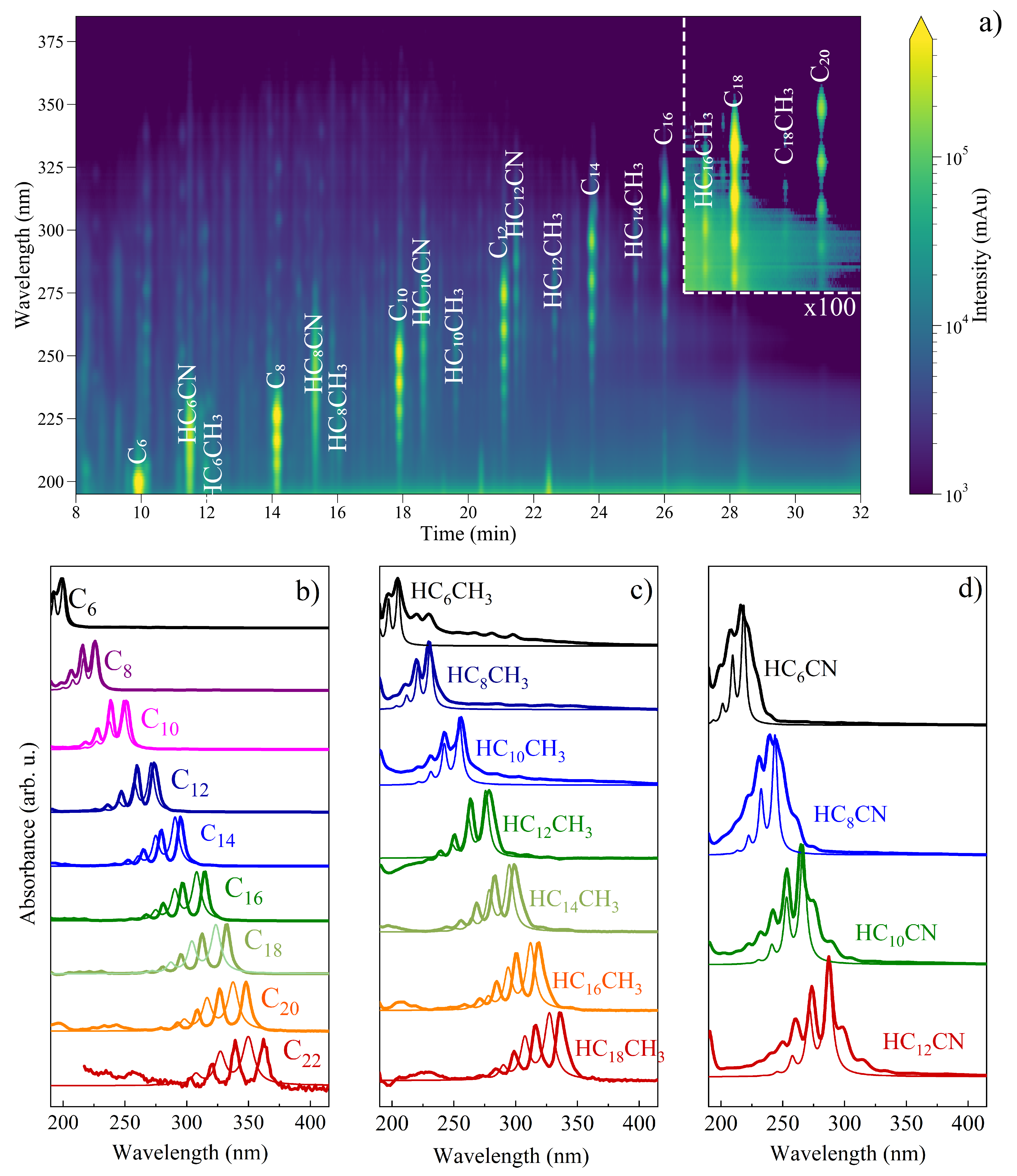}
    \caption{\textbf{a)} 2D graph of the absorbance as a function of the wavelength and time extracted from DAD coupled with HPLC system. Normalized experimental (thick lines) and simulated (thin lines) UV-Vis spectra of \textbf{b)} H-polyynes, \textbf{c)} CH$_3$-polyynes and \textbf{d)} CN-polyynes obtained after ablation of graphite target in acetonitrile.}
\end{figure*}
\\
Polyynes of same size and different terminations have been investigated by Raman spectroscopy. By HPLC, we separated and collected the species with 8 sp-carbon atoms, namely C$_8$, H-C$_8$-CH$_3$ and H-C$_8$-CN. We used this series as a benchmark due to its high concentration and stability, that allows us to efficiently separate them. The low concentration of polyynes ($\lesssim10^{-4}$ M) and the lower Raman activity of short polyynes than the longer ones resulted in a too weak Raman signal which needed to be enhanced by SERS technique \cite{Tabata2006}. Hence, we added silver colloids, to perform SERS in liquids of the three species as shown in \textbf{Figure 4}. We notice two main broad bands below and above 2000 cm$^{-1}$ in all three measurements. The band below 2000 cm$^{-1}$ usually appears in SERS spectra of polyynes, as reported in many works \cite{Lucotti2006,Tabata2006,Lucotti2009}, whereas the band at higher wavenumber corresponds to the ECC mode of sp-carbon chains, consisting of a collective vibrations of the CC bonds, specifically a C$\equiv$C stretching and C$-$C shrinking mode \cite{Agarwal2013}. Concerning the ECC mode, we observe in \textbf{Figure 4} a nearly symmetric peak at 2079 cm$^{-1}$ in C$_8$, a band centred at 2073 cm$^{-1}$ in H-C$_8$-CH$_3$ and a doublet in H-C$_8$-CN, whose most intense peak is located at 2067 cm$^{-1}$ and a secondary one at 2093 cm$^{-1}$. Such differences in the shape and position of ECC band are due to the modulation of polyynes conjugation and possibly to the preferential interaction of silver nanoparticles with the different ends of the chains \cite{Lucotti2006}. In fact, methyl-capped and cyano-capped polyynes present at one end different end-groups which contribute to an increase in the $\pi$-conjugation. CH$_3$ group can give charge to the sp-chain due to an hyperconjugation effect \cite{Castiglioni1989,Castiglioni1986,Jona1981}, while the CN group take part to the $\pi$-electron delocalization due to the presence of sp orbital of N atom. This explains the red shift from C$_8$ (2079 cm$^{-1}$), to H-C$_8$-CH$_3$ (2073 cm$^{-1}$) to H-C$_8$-CN (2067 cm$^{-1}$) \cite{Agarwal2013}. The general trend of the relative positions of these peaks is also shown by simulated Raman spectra of single molecule without Ag nanoparticles reported in \textbf{Figure S.1a} in the Supporting Information. A mismatch between simulated and experimental peaks positions is observed due to the interaction with silver colloids that is not included in our calculations \cite{Lucotti2006,Tabata2006,Lucotti2009}. However, the unpredicted anomalous double-peaked band in SERS spectra of H-C$_8$-CN is probably due to the interaction of silver nanoparticles with two possible adsorption sites, namely H and CN \cite{Tabata2006,extinctioncoeff}. We do not observe the same splitting effect in H-C$_8$-CH$_3$ because we expect the methyl-group to contribute only slightly in the collective vibrational mode due to its sp$^3$ nature (see \textbf{Figure S.1b} in the Supporting Information). The aforementioned asymmetry of ECC band can be the indication of a different interaction with silver nanoparticles in possible different site of the molecule which affects the ECC more to a minor extent. In this picture, the symmetric band of C$_8$ can be related to the presence of two identical terminations as interaction sites with Ag nanoparticles. Even though further computational calculations with suitable models should be developed to analyse more deeply the origin of the experimental SERS spectra, SERS appears to be sensitive to the chain termination showing its potential in polyynes investigation.
\begin{figure}[th]
    \centering
    \includegraphics[width=0.4\paperwidth]{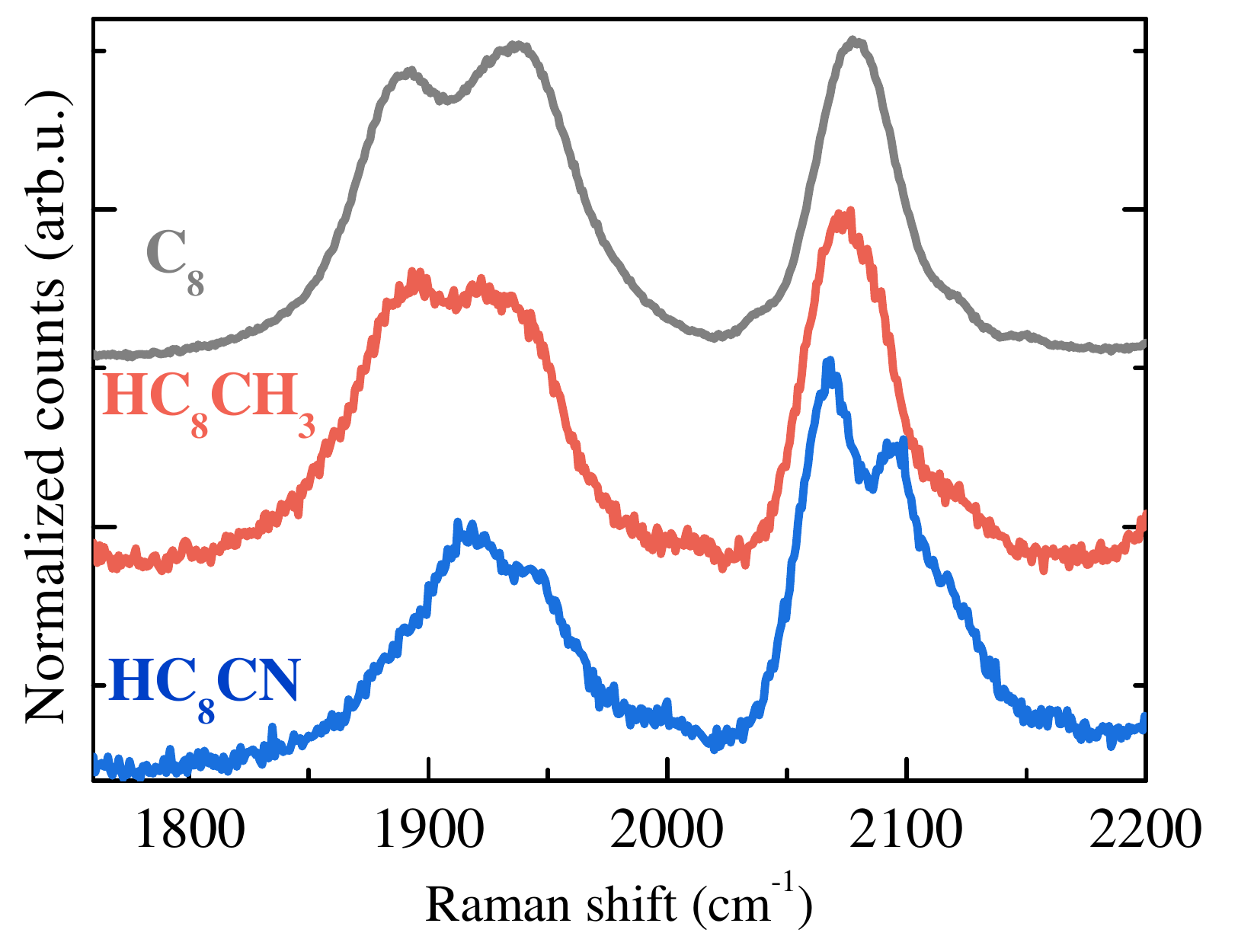}
    \caption{Normalized experimental SERS spectra of liquid solutions of size-selected polyynes with four triple bonds H–(C$\equiv$C)$_4$– and different end-cap (-H/-CH$_3$/-CN).}
\end{figure}
\subsection{Stability of polyynes}
\begin{figure*}[!hb]
    \centering
    \includegraphics[width=0.65\paperwidth]{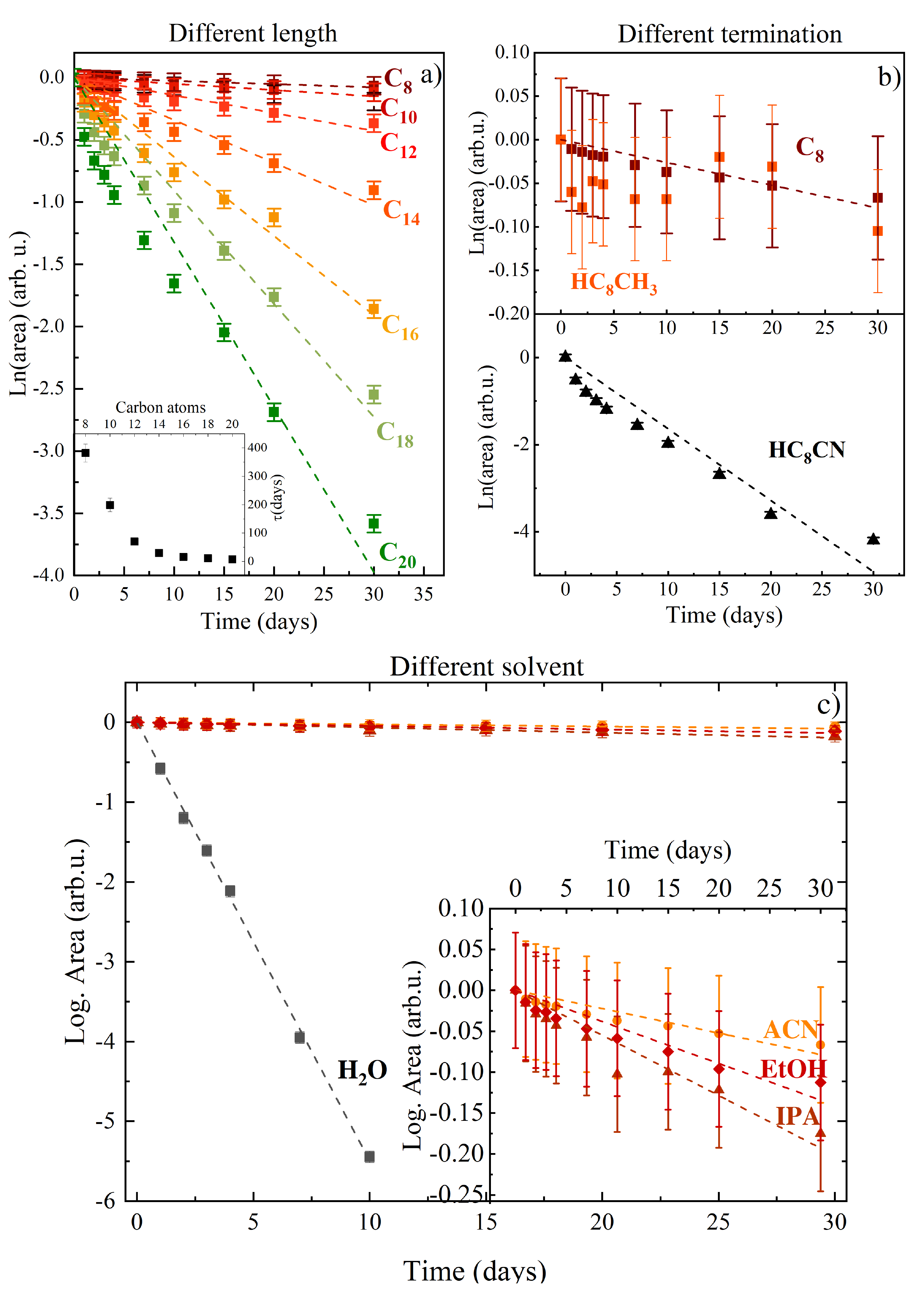}
    \caption{Reduction in time of polyynes chromatographic peak area under specific parameters. \textbf{a)} Effect of the chain length from n=8 to 20. Inset: decay time constant ($\tau$) for each polyynes. \textbf{b)} Effect of the terminations: -H, -CH$_3$ and -CN. \textbf{c)} Effect of the solvents on C$_8$ as water, acetonitrile, isopropanol, ethanol. Inset: Zoom on polyynes solutions in organic solvents. The decay time constants are reported in \textbf{Table S.4} in the Supporting Information.}
\end{figure*} 
We investigated the evolution in time of polyynes as a function of their size and termination. It is known that the degradation of sp-carbon chains is mainly due to the exposure to oxidising agents and to crosslinking interactions \cite{Casari2004,Springborg1992}. Here, we have been studying the effect of some factors on polyynes stability, e.g. chain lengths, terminations and solvents. We are interested in extracting the decay time constants monitoring each solution of polyynes mixture for 30 days by periodic HPLC analysis, keeping them in closed vials and at room conditions. From each measurement, we estimated the amount of selected polyynes at specific times by their chromatographic peak area (A$_t$) and we compared it with the area of the as-prepared solution (A$_0$).The experimental data and the corresponding error bar of 5\%, due to the HPLC apparatus, are presented in \textbf{Figure 5}. In some cases, the temporal evolution of the $\ln{(A_t/A_0)}$ is well described by a linear fitting, from which we calculated the characteristic decay time constants, $\tau$. We show in \textbf{Figure 5a} the area reduction of three size-selected H-polyynes synthesized in acetonitrile, revealing that shorter polyynes are more stable than longer ones, as highlighted in the inset from values of $\tau$ as a function of the number of carbon atoms, as previously discussed by Cataldo\cite{Cataldo2007}. In fact, the shortest polyyne herein reported, i.e. C$_8$, has a decay time constant of 382 days, which is 50 times slower than the longest one (7.6 days), i.e. C$_{20}$, and 4 times greater than C$_{14}$ (29 days), see \textbf{Table S.4} in the Supporting Information. In \textbf{Figure 5b}, for a fixed length of 8 sp-carbon atoms, we compare the different terminations in acetonitrile solution: hydrogen-terminated polyyne (C$_8$), methyl (HC$_8$CH$_3$) and cyano (HC$_8$CN) as terminating groups. HC$_8$CH$_3$ displays a similar evolution in time to that of C$_8$ but with more fluctuations, so the linear fitting could not be performed. We can justify this behaviour considering the steric hindrance of the methyl-group which may reduce the probability of crosslinking reactions between sp-carbon chains and the low concentration of HC$_8$CH$_3$ present in the solution which explains the fluctuations. Furthermore, HC$_8$CN has a decay time constant of 6 days, which is two order of magnitude lower than that of hydrogen-capped polyynes, as already observed by Cataldo\cite{Cataldo2007}. Specifically, this happens because cyanopolyynes, being more reactive than hydrogen-ended ones, oxidise first.\\ 
To investigate the effect of the chemical environment on polyynes stability, we report in \textbf{Figure 5c} the temporal evolution of the chromatographic peak area of the most stable size-selected polyyne C$_8$ in water and organic solvents. Polyynes are not very stable in water, due to their low solubility and non-polarity, as previously discussed. In fact, C$_8$, whose $\tau$ in water is of about 1.8 days, was not anymore detected by HPLC after 10 days. Instead, its decay time constants in all the organic solvents as, for example, acetonitrile (382 days), isopropanol (156 days) and ethanol (224 days), are two order of magnitude higher than the case of water (i.e. 1.8 days) and exhibit only small differences during the analysis period. In fact, the chromatographic peak area of C$_8$ in ACN is decreased by 6\% after 30 days, whereas for EtOH and IPA, the reduction is respectively up to 11\% and 16\%. These data can be explained considering the increased solubility of polyynes in organic solvents as compared with water, due to their lower polarity. Furthermore, ACN, EtOH and IPA have higher $\chi_p$, which causes less by-products that can interact with polyynes and degrade them. Considering the liquid environments investigated, the fact that acetonitrile ensures polyynes stability better than the others can be linked to its minimal value of polarity and the slightest quantity of oxygen dissolved in the solvent \cite{Bebahani2002} (see Ostwald coefficients in \textbf{Table S.2} in the Supporting Information), responsible of the oxidation of the chains \cite{Cataldo2007}. EtOH ensures a slightly greater stability than IPA presumably because, under the same quantity of oxygen present per volume of solvent, the first has a lower $\chi_p$. Increasing the length of H-polyynes, we note the same trend with respect to the solvents. Regarding the methyl-capped polyynes, we notice the irregular behaviour with fluctuations in the order of the experimental error, already described in \textbf{Figure 5b}, not only in ACN but also in IPA and EtOH. In conclusion, the obtained results showed that organic solvents improved polyynes stability compared to water, confirming the improved stability observed when acetonitrile is added to aqueous solution of polyynes \cite{Peggiani2020}.
\section{Conclusions}
We have showed the effect of different solvents on the efficiency of polyynes formation and their stability. Specifically, we have considered water, acetonitrile, methanol, ethanol and isopropanol. H-polyynes up to C$_{22}$ have been separated by HPLC in all the solvents here investigated except for water. The presence of cyano- and methyl- polyynes has been observed by UV-Vis spectra also supported by the comparison with literature and TD-DFT simulations. SERS spectra of isolated size-selected polyynes with the three terminations C$_8$, HC$_8$CH$_3$ and HC$_8$CN have brought out differences in the position and shape of the SERS bands in the sp-carbon range. In addition, we have investigated the evolution in time of the polyynes regarding the impact of the chain length, terminations, and solvents. In conclusion, the degradation in time of short hydrogen-capped polyynes has been well described by a decaying exponential and acetonitrile has provided different advantages: the higher concentration of sp-carbon chains during the laser ablation of graphite, the coexistence of three series of polyynes and the larger stability for H-polyynes with respect to the other solvents investigated. The possibility to exploit participating solvents in PLAL of graphite allows the synthesis of polyynes with different termination opening to the modulation of the properties by selection of size and type of endgroup.

\section*{Conflicts of interest}
There are no conflicts to declare.

\section*{Acknowledgements}
Authors acknowledge funding from the European Research Council (ERC) under the European Union’s Horizon 2020 research and innovation program ERC-Consolidator Grant (ERC CoG 2016 EspLORE grant agreement No. 724610, website: www.esplore.polimi.it).

\bibliography{references}
\bibliographystyle{unsrt}



\section*{Supplementary material}

# Supporting information

## Solvent-dependent termination, size and stability in polyynes synthesis by laser ablation in liquids

Sonia Peggiani, Pietro Marabotti, Riccardo Alberto Lotti, Anna Facibeni, Patrick Serafini, Alberto Milani, Valeria Russo, Andrea Li Bassi and Carlo S. Casari

*Department of Energy, Politecnico di Milano, Via Ponzio 34/3, 20133, Milano, Italy*

**Table S.1** State of art of different polyynes obtained by physical methods in terms of length and terminations. The references are listed below.

| n. of carbon atoms (end-cap) | $H_2O$ | ACN | MeOH | EtOH | 1-propanolo | 1-butanolo | t-butyl alcohol | hexane | n-hexane | c-hexane | n-heptane | n-hoctane | decane | toluene | benzene | decalin | TEOS |
|---|---|---|---|---|---|---|---|---|---|---|---|---|---|---|---|---|---|
| 6 (H) | 1 | 2 | | | | | | | | | | | | | | | |
| 6 ($CH_3$) | | | | | | | | | | | | | | | | | |
| 6 (CN) | | 3 | | | | | | | | | | | | | | | |
| 8 (H) | 1, 2, 4-8 | 2, 3, 7, 9 | 7, 10-12 | 10, 13-15 | 10 | 10 | 10 | 16-18 | 5, 9, 11, 12, 19 | 7, 9 | 19 | 19 | 18 | 12, 20, 21 | 21, 22 | 23, 24 | 25 |
| 8($CH_3$) | | | | | | | | | | | | | | | | | |
| 8(CN) | | 3 | | | | | | | | | | | | | | | |
| 10(H) | 1, 4, 5, 7, 8 | 2, 3, 7, 9 | 7, 10-12 7, 10-12 | 10, 13-15 | 10 | 10 | 10 | 16-18 | 5, 9, 11, 12, 19 | 7, 9 | 19 | 19 | 18 | 12, 20, 21 | 21, 22 | 23, 24 | 25 |
| 10($CH_3$) | | | | | | | | | | | | | | | | | |
| 10(CN) | | 3 | | | | | | | | | | | | | | | |
| 12 (H) | 7 | 2, 7, 9 | 10, 11 | 10, 13-15 | 10 | 10 | 10 | 16, 17 | 5, 9, 11, 12, 19 | 9 | 19 | 19 | | 12, 20, 21 | 21, 22 | 23, 24 | 25 |
| 12($CH_3$) | | | | | | | | | | | | | | | | | |
| 12(CN) | | 3 | | | | | | | | | | | | | | | |
| 14 (H) | | 2, 7, 9 | 10, 11 | 10, 13-15 | 10 | 10 | 10 | 16, 17 | 5, 9, 11, 19 | 9 | 19 | 19 | | 12, 20, 21 | 21, 22 | 23, 24 | 25 |
| 14($CH_3$) | | 20 | | | | | | | | | | | | | | | |
| 16(H) | | 2, 9 | 10, 11 | 10, 13-15 | 10 | 10 | 10 | 16, 17 | 5, 9, 11 | 9 | | 19 | | 12, 20, 21 | 21 | 23, 24 | 25 |
| 16($CH_3$) | | | | | | | | | | | | | | | | | |
| 18(H) | | | 11 | | | | | 16, 17 | 5, 11 | | | | | 20, 21 | | 23, 24 | |
| 18($CH_3$) | | | | | | | | | | | | | | | | | |
| 20(H) | | | 11 | | | | | 16, 17 | 5, 11 | | | | | | | 23, 24 | |
| 22(H) | | | | | | | | 17 | 11 | | | | | | | 23, 24 | |

**Table S.2** Molar concentration (mol/L) of H-polyynes of different size in different solvents. Polarity values indicated with "p" are taken from the Handbook of organic solvents properties [26] and "l" for Oswald coefficient, defined as the ratio of concentrations of the gas in the liquid and gas phases.

|  | ACN<br>p=46<br>l=0.00083[27] | IPA<br>p=54.6<br>l=0.2463[28] | EtOH<br>p=65.4<br>l=0.2417[28] | MeOH<br>p=76.2<br>l=0.2476[28] | $H_2O$<br>p=100<br>l=0.031[29] |
|---|---|---|---|---|---|
| $C_8$ | $(1.42\pm0.07)\times10^{-4}$ | $(1.09\pm0.02)\times10^{-4}$ | $(1.07\pm0.01)\times10^{-4}$ | $(7.50\pm0.05)\times10^{-5}$ | $(3.13\pm0.02)\times10^{-6}$ |
| $C_{10}$ | $(7.3\pm0.4)\times10^{-5}$ | $(4.57\pm0.08)\times10^{-5}$ | $(4.07\pm0.06)\times10^{-5}$ | $(2.72\pm0.03)\times10^{-5}$ | 0 |
| $C_{12}$ | $(2.6\pm0.1)\times10^{-5}$ | $(1.8\pm0.1)\times10^{-5}$ | $(9.3\pm0.6)\times10^{-6}$ | $(1.17\pm0.02)\times10^{-5}$ | 0 |
| $C_{14}$ | $(7.8\pm0.4)\times10^{-6}$ | $(1.1\pm0.4)\times10^{-5}$ | $(8.0\pm0.3)\times10^{-6}$ | $(4.6\pm0.2)\times10^{-6}$ | 0 |
| $C_{16}$ | $(6.5\pm0.3)\times10^{-6}$ | $(3.4\pm0.1)\times10^{-6}$ | $(3.0\pm0.1)\times10^{-6}$ | $(1.6\pm0.1)\times10^{-6}$ | 0 |

**Table S.3** Polyynes ended by –H/ –CH$_3$/ –CN obtained after ablation of graphite target in acetonitrile with the corresponding times on the chromatogram and the positions of the experimental, simulated and literature UV-Vis absorption peaks. References are listed below.

| n. of C atoms<br>(end-cap) | $t_R$(min) | Wavelength(nm) | | |
|---|---|---|---|---|
|  |  | *Experimental* | *Simulated* | *Literature* |
| 6(H) | 9.919 | 198 | 200 | 199 [1] |
| 6(CN) | 11.484 | 216  208  200 | 218  209  201 | 215.6  207.2  198.9 [3] |
| 6(CH$_3$) | 11.98 | 205  197 | 205 | / |
| 8(H) | 14.149 | 226  216  207 | 226  216  208 | 226  216  206 [23] |
| 8(CN) | 15.323 | 239/244.5  231  222 | 244  232  222 | 239/244  231  222 [3] |
| 8(CH$_3$) | 16.058 | 230  220  211 | 231  221  212 | / |
| 10(H) | 17.89 | 251  238  228 | 249  237  227 | 251  239  227 [23] |
| 10(CN) | 18.627 | 265  253  242 | 266  253  241 | 264.5  253.3  242.0 [3] |
| 10(CH$_3$) | 19.623 | 256  242  231 | 254  242  231 | 257  243  232 [30] |
| 12(H) | 21.092 | 273  260  247 | 271  257  245 | 275  260  247 [23] |
| 12(CN) | 21.468 | 287  273  261 | 287  271  258 | 287.4  273.7  261.1 [3] |
| 12(CH$_3$) | 22.64 | 278  263  251 | 276  261  249 | 279  264  251 [20] |
| 14(H) | 23.777 | 295  280  265 | 291  275  261 | 296  280  267 [23] |
| 14(CH$_3$) | 25.120 | 299  283  268 | 295  279  264 | 302  285  269 [20] |
| 16(H) | 26.004 | 315  296  281 | 308  290  275 | 316  298  281 [23] |
| 16(CH$_3$) | 27.251 | 319  301  285 | 312  294  278 | / |
| 18(H) | 28.045 | 333  313  295 | 324  304  287 | 334  314  295 [23] |
| 18(CH$_3$) | 29.691 | 335  316  299 | 327  307  290 | / |
| 20(H) | 30.795 | 348  326  309 | 337  316  298 | 350  328  310 [23] |
| 22(H) | 34.288 | 362  339  321 | 350  327  308 | 364  341  321 [23] |

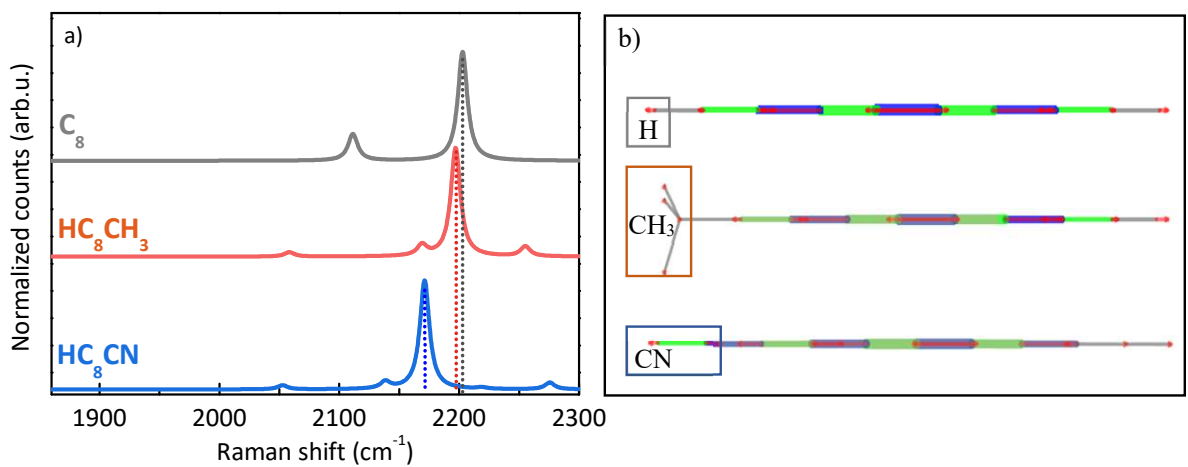

**Figure S.1** a) Normalized simulated Raman spectra of hydrogen-, methyl-, cyano-capped polyynes with four triple bonds –(C≡C)$_4$–. b) The collective vibrational mode of CC bond related to the same molecules. Simulated Raman spectra and relative vibrational mode have been computed by PBE0/cc-pVTZ calculations (see Section 2).

**Table S.4** Decay time constant and corresponding $R^2$ of the fit of polyynes changing length, termination and solvent.

| Length | τ(days) | $R^2$ |
|---|---|---|
| $C_8$ | -382.381 ± 31.578 | 0.93575 |
| $C_{10}$ | -198.860 ± 24.259 | 0.86876 |
| $C_{12}$ | -70.187 ± 5.949 | 0.93252 |
| $C_{14}$ | -29.223 ± 2.206 | 0.94579 |
| $C_{16}$ | -15.708 ± 0.837 | 0.97235 |
| $C_{18}$ | -11.002 ± 0.600 | 0.97104 |
| $C_{20}$ | -7.553 ± 0.464 | 0.96353 |

| Termination | τ(days) | $R^2$ |
|---|---|---|
| -H | -382.381 ± 31.578 | 0.93575 |
| -CN | -6.105 ± 0.413 | 0.95606 |

| Solvent | τ(days) | $R^2$ |
|---|---|---|
| $H_2O$ | -1.8191 ± 0.018 | 0.99929 |
| ACN | -382.381 ± 31.578 | 0.93575 |
| IPA | -155.438 ± 10.500 | 0.95617 |
| EtOH | -223.464 ± 17.020 | 0.94487 |

# References of Table S.1 and S.3


\end{document}